\documentclass[aps,pre,
superscriptaddress,groupedaddress,showpacs]{revtex4}
\usepackage{graphicx,epsf}
\usepackage{xcolor}
\usepackage{psfrag}
\usepackage[english]{babel}
\begin{document}
%----------------------------------------------------------------------------
%                              TITLE
%----------------------------------------------------------------------------
\title{The size and shape of snowflake star polymers in dilute solutions:\\ analytical and numerical approaches}
%----------------------------------------------------------------------------
%                              AUTHORS revtex4-style
%----------------------------------------------------------------------------
\author{K. Haydukivska}
\affiliation
{Institute of Physics, University of Silesia, 41-500 Chorz\'ow, Poland}
\affiliation{Institute for Condensed
Matter Physics of the National Academy of Sciences of Ukraine,\\
79011 Lviv, Ukraine}
\author{V. Blavatska}
\affiliation{Institute for Condensed
Matter Physics of the National Academy of Sciences of Ukraine,\\
79011 Lviv, Ukraine}
\affiliation{Dioscuri Centre for Physics and Chemistry of Bacteria,
Institute of Physical Chemistry, Polish Academy of Sciences, 01-224 Warsaw, Poland}
\author{Jaros{\l}aw Paturej}
\affiliation
{Institute of Physics, University of Silesia, 41-500 Chorz\'ow, Poland}
\affiliation{Leibniz-Institut f\"ur Polymerforschung Dresden e.V., 01069 Dresden,
Germany}
\email{jaroslaw.paturej@us.edu.pl}

%----------------------------------------------------------------------------
%                             ABSTRACT
%----------------------------------------------------------------------------
\begin{abstract}
We investigate the conformational properties of a multi-branched polymer structure with a dendrimer-like topology, known as a snowflake polymer. This polymer is characterized by two parameters: $f_s$, which represents the functionality of the central star-like core, and $f$, which represents the functionality of the side branching points.
To analyze the conformational properties, we have employed various approaches, including analytical methods based on direct polymer renormalization and the Wei's approach as well as numerical molecular dynamics simulations. These methods have allowed us to estimate a size and shape characteristics of the snowflake polymer as functions of $f$ and $f_s$.
Our findings consistently demonstrate the effective compactification of the typical polymer conformation as the number of branching points increases. 
Overall, our study provides valuable insights into the conformational behavior of the snowflake polymer and highlights the impact of branching parameters on its overall compactness. 
\end{abstract}
\pacs{36.20.-r, 36.20.Ey, 64.60.ae}
\date{\today}
\maketitle

\section{Introduction}

Progress in polymer chemistry has paved the way for the creation of intricate polymer structures, including hyperbranched structures \cite{Duro2015, England2010, Higashihara11, Zheng15, polymeropoulos2017}. It has been observed that the presence of multiple branching points within a single macromolecule significantly alters the rheological properties of their melts \cite{McLeish98}. Additionally, these structures are anticipated to have a significant impact on targeted drug delivery \cite{Duro2015, Saying14}, encapsulation of dyes \cite{Higashihara11,Yu14}, and purification of proteins \cite{England2010,Carter06}.

\begin{figure}[b!]
\begin{center}
\includegraphics[width=53mm]{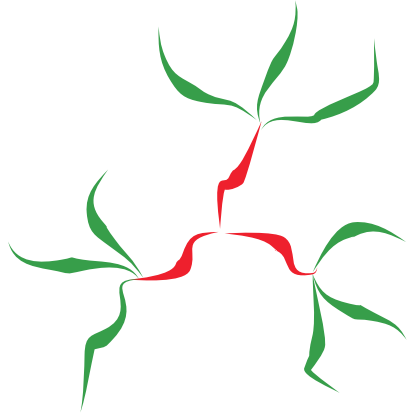}
\caption{ \label{fig:1} Schematic presentation of a snowflake polymer with functionality of the central core $f_s=3$ (red) and functionality of the side branching points $f=4$ (green).}
\end{center}
\end{figure}

The simplest case of a multi-branched structure, characterized by more than one branching point, is known as a pom-pom polymer \cite{Zimm49}. This polymer consists of a linear backbone with two branching points, denoted as $f_1$ and $f_2$, located at both ends. The properties of these molecules have been extensively investigated in both melts \cite{McLeish98, Graham01, Ruymbeke07, Chen11} and solutions \cite{Zimm49, 1996, Kalyuzhnyi20, Haydukivska22}. In comparison to single-branched star-like topologies, the presence of multiple long-chain branches in pom-pom polymers leads to a decrease in viscosity and the occurrence of strain-hardening phenomena during uniaxial extensional flow \cite{McLeish98}.

Bottlebrush polymers serve as significant examples of complex multibranched structures. In these polymers, a collection of $f_c$ side chains is regularly attached to a linear polymer backbone at $n$ branching points \cite{Sheiko}. 
Th variation of bottlebrush architecture  provides unique features in dilute solutions and melts
such as architecture-dependent physical properties including
structure \cite{paturej1,paturej2} and rheology \cite{paturej3}.  
In particular the attachment of side chains to the backbone induces a strong spatial correlation among the monomers of the backbone,  due to the steric repulsion exerted by the side chains. Consequently, bottlebrush conformations are  characterized by enhanced stiffness of the backbone as compared to linear chains \cite{Terao,Kawaguchi}.

 Dendritic macromolecules represent another category of hyperbranched topologies, characterized by successive branching units. Dendrimers  are composed of a central core, branched units called "generations," and terminal functional groups. 
  The structure of a dendrimer resembles that of a tree, with branches extending outwards from the core. 
 %These structures bear resemblance to a Cayley tree with a star-like cascade topology.
In 1978, Fritz Vogtle and co-workers successfully synthesized a complex polymer structure, which they named "cascade molecules" \cite{Vogtle}. Later, in 1985, D.A. Tomalia {\it et al.} utilized the term "dendrimers" to describe similar structures that they synthesized \cite{Tomalia}.
  Dendrimers possess an initiator core (the first generation) which may be presented as star consisting of $f_s$ branches, and $G\geq 1$ successive layers-generations composed of repeating units with branching points of functionalities $f$. On Fig. \ref{fig:1} we schematically present the structure corresponding to dendrimer topology of the first generation $G=1$. Note that $G=0$ generation corresponds to single-branch star-like topology.
  A comprehensive and recent review discussing the properties and potential applications of dendrimers across different fields of research, technology and treatment can be found in Ref.~\cite{Mathur10}.  A recent study by Liu et al. \cite{Liu21} introduced a novel category of dendronized arm snowflake polymers with functional cores. These polymers were synthesized as an improved alternative to high-generation dendrimers with high molecular weight, while still retaining their essential functional characteristics, including high encapsulating efficiency and biocompatibility.  
In the present work, we  consider the simplified model of a snowflake-shaped polymer  with functionality $f_s$ of central core and denronized arms functionality $f$ (cf. Fig~\ref{fig:1}).
  
   The study of the conformation properties of macromolecules in dilute solutions is at the heart of polymer science \cite{wang}. Essentially all physical properties of polymers are manifestations of the underlying polymer conformations or otherwise significantly impacted by the conformation properties.
At  low polymer concentrations, the intermolecular interactions between them can be disregarded, and the measurable quantities such as gyration radius ($R_g$) or hydrodynamic radius ($R_H$) are primarily determined by the individual molecular topology of polymers \cite{burchard}. The conformational properties of macromolecules in dilute solution regime allow the relatively simple analytical treatment  \cite{desCloiseaux, Zimm49,Douglas84}, yielding results that align well with experimental 
 data \cite{Douglas84,Paturej22} and numerical simulations \cite{Kaluzhniy22,Paturej22}.
   The central point of these analytical approaches is the independence of essential size and size-related characteristics from specific details of the chemical structure of macromolecules. This enables the study of a diverse range of chemically distinct yet topologically similar molecules. \cite{desCloiseaux}. 
The typical example of such structural quantity is the so-called size ratio $g_{{\rm complex}}$ which is defined as a ratio of the radius of gyration $\langle R_g^2 \rangle_{\rm complex}$ of a polymer witch complex architecture  to the radius of gyration $\langle R_g^2 \rangle_{\rm chain}$
of a linear chain of the the same total molecular weight \cite{Zimm49}: 
\begin{equation}
g_{{\rm complex}}=\frac{\langle R_g^2 \rangle_{\rm complex}}{\langle R_g^2 \rangle_{\rm linear}}. \label{gratio}
\end{equation}
The quantity $g_{{\rm complex}}$ is often used to characterize the impact of the complex polymer topology  on its effective contraction/elongation in solvent. Here and below the symbol $\langle (\ldots) \rangle$ denotes averaging over an ensemble of possible polymer conformations. The quantity $g_{{\rm complex}}$ can be estimated exactly for ideal Gaussian polymers without taking into account the excluded volume interactions. In particular, for the case of a single-branched  star-like structure, corresponding to snowflake with side functionalities $f=1$, the size ratio is given by \cite{Zimm49}:  
\begin{equation}
g_{{\rm star}}=\frac{3f_s-2}{f_s^2}. \label{gstar}
\end{equation}
The size ratio for pom-pom topology corresponding to a snowflake polymer with $f_s=2$ and with functionalities of two side chains equal to $f_1$ and $f_2$ is given by 
 equation \cite{Zimm49,Radke96}: 
\begin{equation}
g_{{\rm pom-pom}}=\frac{3(f_1^2+f_2^2)+4(f_1+f_2)+12f_1f_2+1}{(f_1+f_2+1)^2}. \label{gpom}
\end{equation}
  The analytical estimates for the Gaussian 
bottlebrush polymers containing $n$ branching points of functionality $f$ can be found in Ref.~\cite{Nakamura}, whereas for an ideal dendrimer polymers were obtained in Refs.~\cite{Boris96,Ferla97}. A set of numerical estimates for the radius of gyration of dendrimer polymers of different generations $G$ can be found in Refs.~\cite{Timoshenko02,Sheng08,Ganazzoli00,Ganazzoli02,Mansfield00,Tande01}.

The effective impact of macromolecule topology on shape in a solvent  can be also characterised  by  the asphericity $A_d$ which is  defined as \cite{Aronovitz,Rudnick86}: 
\begin{equation}
\langle A_d\rangle=\frac{1}{d(d-1)}\left\langle\frac{{\rm Tr}\, \hat{\bf S}^2 }{({\rm Tr}\, {\bf S})^2} \right\rangle.\label{Ad}
\end{equation}
In the above equation $ {\bf S}$ in the gyration tensor and $\hat{ {\bf S}}= {\bf S}-\overline{\mu}{\bf I}$ with $\overline{\mu}$ denotes an average eigenvalue whereas ${\bf I}$ is a unity matrix. 
The 
quantity $A_d$ describes the deviation of a polymer shape from a spherical one (with $A_d=0$) and reaches its maximum value of 1 for completely stretched, rod-like conformation.
To get experimentally observed average value $\langle A_d \rangle$ one has to perform averaging  
 over an ensemble
of all possible polymer configurations. Note that most of  analytical studies \cite{Aronovitz,Rudnick86,Blavatska11}  avoid the averaging of the
ratio in Eq.~(\ref{Ad}) due to essential difficulties in calculations and evaluate the quantity 
$\hat{A_d}$
where the numerator and the denominator of Eq.~(\ref{Ad}) are averaged separately. The value for averaged asphericity
$\langle A_d \rangle$ can  be  obtained   for any complex Gaussian polymer architecture applying numerical approach \cite{Ferber15,Blavatska20,haydukivskauniversal}.

The layout of the paper is as follows. We start with analytical description of  snowflake-shaped polymers by adapting the analytical  continuous chain model in Sec.~\ref{II}, followed by the introduction of the system in terms of the numerical bead-spring model (Sec.~\ref{III}) and a graph-based model of the Wei's method in Sec.~\ref{Wei}. The results obtained by three different approaches are compared and discussed in Sec.~\ref{IIII}. We end up with concluding remarks presented in Sec.~\ref{Con}.  

\section{Analytical approach}
\label{II}
\subsection{Continuous chain model}
\label{M}

Following the scheme   developed in Ref. \cite{Edwards}, 
we consider a single polymer chain as a trajectory of length $L$ parameterized with radius vector $\vec{r}(s)$. Any complex topology can be presented as a set of trajectories of the same length.  The hamiltonian for snowflake-like  topology presented in  Fig.~\ref{fig:1} within continuous chain model can be presented as:
\begin{eqnarray}
&&H = \frac{1}{2}\sum_{i=1}^{F}\,\int_0^L ds\,\left(\frac{d\vec{r_i}(s)}{ds}\right)^2\nonumber\\
&&+\frac{u}{2}\sum_{i,j=1}^{F}\int_0^Lds'\int_0^L ds''\,\delta(\vec{r_i}(s')-\vec{r_j}(s'')),\label{H}
\end{eqnarray}
where $F$ denotes the number of trajectories in the structure under consideration $F=\sum_{i=1}^{f_s} f_i$ with $f_s$ being the functionality of core branching points   and $f_i$ functionalities of external branches. The first term in the hamiltonian of Eq.~(\ref{H}) describes the connectivity of each chain and the second term describes the excluded volume interaction with coupling constant $u$.

%The difference between topologies is introduced %in the averaging of observables over ensemble %of different configurations, in the simplest %case the partition function (number of possible %configurations). In the case of one branching %point it can be presented as:
%\begin{eqnarray}
%Z^{star}_{F}=\frac{1}%%%{Z_0}\prod_{i=1}^{F}\int\,D\vec{r}(s)\,\delta(\vec{r_i}(0))\,{\rm e}^{-H},
%\label{Zs}
%\end{eqnarray}
%here  
The partition sum of this model is given by
\begin{eqnarray}
&&Z^{\rm snowflake}_{\{f_i\},f_s}=\frac{1}{Z_0}\prod_{i=1}^{f_s}\prod_{j=2}^{f_i}\,\int\,D\vec{r}(s)\,\delta(\vec{r^i_1}(L))\times\nonumber\\
&&\times\delta(\vec{r^i_1}(0)-\vec{r^i_j}(0))\,{\rm e}^{-H},
\label{ZZ}
\end{eqnarray}
where the products of $\delta$-functions stand for beginning of all trajectories $\vec{r_i}$ at the origin, $Z_0$ is a partition sum of a Gaussian polymer without excluded volume interactions.

Any observables, calculated on the basis of continuous chain model, depend on the chain length $L$ and diverge in asymptotic limit  $L\rightarrow \infty$. These divergences  can be removed within the direct polymer renormalization procedure \cite{desCloiseaux}. The main goal of this procedure is evaluation of the reliable physical values of observables at the so-called fixed points (FPs) of the renormalization group.
It is a very useful  that a set of FPs does not depend on polymer topology, and it is enough to make use of FPs obtained from a simplest topology of a linear chain, which are well known \cite{desCloiseaux}. These FPs can be presented as results of perturbation group series in deviation from the upper critical dimension $\epsilon=4-d$ and read:
\begin{eqnarray}
&& {\rm {Gaussian}}: u^*_{R}=0, \label{FPG}\\
&& { \rm {Pure}}: u^*_{R}=\frac{\epsilon}{8}, \label{FPP}
 \end{eqnarray}
where Eq.~(\ref{FPG}) describes an ideal Gaussian polymer, and Eq.~(\ref{FPP}) provides a coupling constant for model with excluded volume interactions.

\section{Molecular Dynamics}
\label{III}
Numerical simulations were conducted using a three-dimensional, bead-spring coarse-grained model \cite{grest1987}.  Each chain in the snowflake architecture was represented as a sequence of  $N$ beads connceted with each other. The total number of beads in a snowflake polymer including one bead in the center is $f_s f N+1$. The beads are connected into the chains by springs described with the finitely-extensible
nonlinear elastic (FENE) potential:
\begin{equation}
 V^{\mbox{\tiny FENE}}(r)=- 0.5kr_0^2\ln{[1-(r/{r_0})^2]}.
\label{fene}
\end{equation}
The excluded volume interactions are introduced by repulsive part of the Lennard-Jones potential that was shifted and truncated, known as a Weeks-Chandler-Anderson (WCA) interaction:
\begin{equation}
 V^{\mbox{\tiny WCA}}(r) = 4\epsilon_{LJ}\left[
(\sigma_{LJ}/ r)^{12} - (\sigma_{LJ} /r)^6 + 1/4
\right]\theta(2^{1/6}\sigma_{LJ}-r).
\label{wca}
\end{equation}

In the above equation $r$ is the distance between the centers of the beads with diameter $\sigma_{LJ}$, $\epsilon_{LJ}$ is  an energy scale and the constants being $k=30\epsilon_{LJ}/\sigma_{LJ}^2$ and $r_0=1.5\sigma$. In Eq.~(\ref{wca}) we introduced the Heaviside step function defined as $\theta(x)=0$ or 1 for $x<0$ or $x\geq 0$. 
%In total these potentials are known as Kremer-Grest potentia~\cite{grest1986}  $V^{\mbox{\tiny %KG}}(r)=V^{\mbox{\tiny
%FENE}}(r)+V^{\mbox{\tiny WCA}}(r)$.

The simulations were run using the Large-scale Atomic/Molecular Massively Parallel Simulator (LAMMPS) \cite{lammps}. We numerically solved the Newton's equations of motion through the velocity-Verlet algorithm with the iteration step  $\Delta t = 0.005\tau$. We utilized NVT enseble in our simulations and the temperature  $T$ was maintained by presence of the Langevin dumping term with the coefficient
$\zeta=0.5\,m\tau^{-1}$, where $\tau = \sqrt{m\sigma^2/\epsilon}$ is the LJ time unit and $m=1$ is monomer mass. The periodic boundary conditions in all three dimensions for the cubic box were implemented in all simulations.

\begin{figure}
\begin{center}
\includegraphics[scale=0.3]{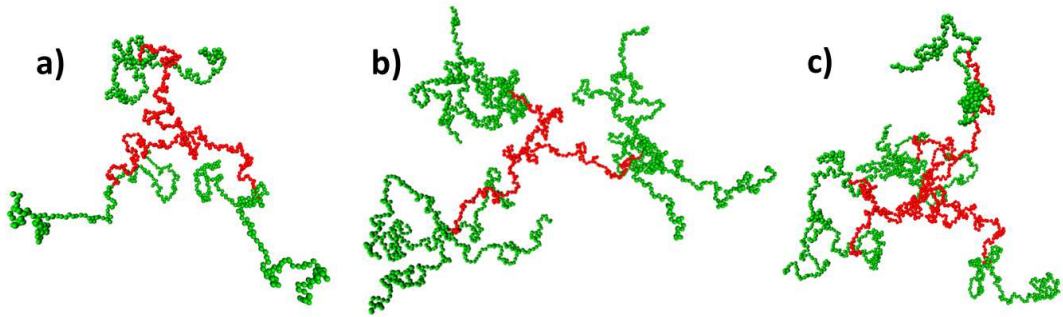}
\caption{ \label{snap} Molecular dynamics snapshots of snowflake-shaped polymers with functionality of the central core $f_s$ (depicted in red) and functionality of the side branching points $f=4$ (in green): a) $f_s=3$ and $f=3$, b) $f_s=3$ and $f=6$ and c) $f_s=6$ and $f=3$.}
\end{center}
\end{figure}

Initial conformations of polymers were generated  using self-avoiding walk technique on a simple cubic lattice by implementation of the pivot algorithm \cite{Madras88} with $20 f_s f N$ pivot steps. By incorporating the pivot step into molecular dynamic simulations, significant changes in the configuration of the molecules can be achieved. This implementation enables the initiation of simulations from a more compact state, resulting in reduced computational time.

Each simulation box contained $27$ molecules. We did not accounted for inter-molecular interactions between the molecules in order to capture dilute solution conditions. The simulations were conducted for $1.4\cdot10^8$ and $2\cdot10^8$ integration steps, respectively, for systems containing $N=50$ and $N=100$ beads. The calculation of the gyration radius involved a minimum of $10^7 \tau$ steps, with the averaging process commencing after three autocorrelation times had elapsed.

The branched architectures of snowflake polymers with $f_s=3$ and $f$ from $3$ to $6$ were investigated   for the arm lengths composed of $N=50$ and $N=100$ beads. The values of the radius of gyration of linear polymers were calculated using a fitting function based on the simulations results conducted for the chain lengths between $100$ and $600$ beads. In that range of $N$ we observed  a proper scaling behaviour of linear polymer in good solvent and the final fitting was carried out with scaling exponents $\nu=$ and $\Delta=$ taken from Ref.~\cite{Clisby10}.

\section{Numerical approach: Wei's method}
\label{Wei}

Any complex polymer structure 
can be described as a mathematical graph (network), where the individual monomers are presented as vertices, and the chemical bonds between monomers are considered as links between them. The chemical functionalities of monomers  are then equal to degrees of corresponding vertices.
The Wei's method \cite{Wei} is utilized  to estimate the size and shape properties of polymer network of any topology, if  the
Kirchhoff matrix and its eigenvalues are defined. 
For the polymer architecture of total number of $M$ monomers, Kirchhoff $M\times M$ matrix ${\bf K}$ is defined as follows. 
Its diagonal elements $K_{ii}$ are equal
to the degree of vertex $i$, whereas the   non-diagonal elements $K_{ij}$ are equal to $-1$ when the vertices $i$ and $j$ are adjacent and $0$ otherwise. 

 Let $\lambda_2,\ldots,\lambda_M$ be $(M-1)$ non-zero eigenvalues of the $M\times M$
Kirchhoff matrix 
\begin{equation}
    {\bf K}{\bf Q}_i =\lambda_i {\bf Q}_i,\,\, \,\,\,\,\,i=1\ldots M 
    \end{equation}
 ($\lambda_1$ is always $0$). 
 The $g$-ratio of the radius of gyration of a topologically complex network to that of a linear chain with the same molecular weight reads:
\begin{equation}
g=\frac{\sum_{j=2}^{M}1/\lambda_j^{{\rm network}}}{\sum_{j=2}^{M}1/\lambda_j^{{\rm chain}}},
\label{gwei}
\end{equation}
where $\lambda_j^{{\rm network}}$ and $\lambda_j^{{\rm chain}}$ are the network and the linear chain   Kirchhoff matrix eigenvalues, respectively.
The
asphericity  in $d$ dimensions is  given by \cite{Wei,Ferber15}:
\begin{equation}
\langle A_d \rangle =\frac{d(d+2)}{2}\int_0^{\infty} {\rm d} y \sum_{j=2}^{M}\frac{y^3}{(\lambda_j+y^2)^2}\left[ \prod_{k=2}^{M} 
\frac{\lambda_k}{\lambda_k+y^2}\right ]^{d/2}.
\label{awei}\end{equation}

To construct a polymer network on a basis of a snowflake architecture, we consider each link in a graph as a polymer chain with number of monomers (treated as vertices) $n$, and each vertex with degree $k > 1$ as a junction point, so that the resulting graph contains $M = N +L \times l$ vertices.

\section{Results and Discussion}

\label{IIII}
\subsection{Partition function}

\begin{figure}
\begin{center}
\includegraphics[width=73mm]{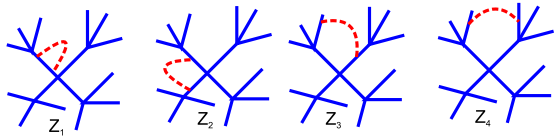}
\caption{ \label{fig:2}Diagrammatic presentation of contributions into the partition function in one-loop approximation. The solid lines are schematic presentations of polymer pathes each of length $L$ and dash line represents a two monomer excluded volume interaction.}
\end{center}
\end{figure}

We start our discussion with evaluating the partition $Z^{\rm snowflake}$ function of a snowflake-shaped polymer:
\begin{eqnarray}
&&Z^{\rm snowflake}_{\{f_i\},f_s}=\frac{1}{Z_0}\prod_{i=1}^{f_s}\prod_{j=2}^{f_i}\,\int\,D\vec{r}(s)\,\delta(\vec{r^i_1}(L))\times\nonumber\\
&&\times\delta(\vec{r^i_j}(0)-\vec{r^i_1}(0))\exp\left({-\frac{1}{2}\sum_{i=1}^{F}\,\int_0^L ds\,\left(\frac{d\vec{r_i}(s)}{ds}\right)^2}-\right.\nonumber\\
&&-\frac{u}{2}\sum_{i,j=1}^{F}\int_0^Lds'\int_0^L ds''\,\delta(\vec{r_i}(s')-\vec{r_j}(s''))\nonumber\\
&&\left.{-\frac{1}{2}\sum_{i=1}^{F}\,\int_0^L ds\,\left(\frac{d\vec{r_i}(s)}{ds}\right)^2}\right)\label{ZZ1}
\end{eqnarray}
The partition function $Z_0$ of a Gaussian chain can be determined from:
\begin{eqnarray}
&& Z_0=\prod_{i=1}^{f_s}\prod_{j=2}^{f_i}\,\int\,D\vec{r}(s)\,\delta(\vec{r^i_1}(L))\delta(\vec{r^i_j}(0)-\vec{r^i_1}(0))\times\nonumber\\
 &&\times{\rm e}^{-\frac{1}{2}\sum_{i=1}^{F}\,\int_0^L d s\,\left(\frac{d\vec{r_i}(s)}{d s}\right)^2}
\end{eqnarray}
The contributions from excluded volume interactions are considered to be much smaller as compared with the contribution of Gaussian elasticity term. Consequently, it is common practice to employ a perturbation theory expansion based on the coupling constant $u$ to evaluate these steric effects.
To calculate the second term in Eq.~(\ref{ZZ1}) we use a Fourier representation for the $\delta$ function:
\begin{eqnarray}
\delta (\vec{r}_i(s')-\vec{r}_j(s'')) =\frac{1}{(2\pi)^{d}} \int {\rm d}\vec{p_u}\, {\rm e}^{\left(-\iota\vec{p_u}(\vec{r}_i(s')-\vec{r}_j(s'')\right)}
\end{eqnarray}
To proceed with calculations, we utilize the diagrammatic technique~\cite{desCloiseaux}. In Fig.~\ref{fig:2} we show different terms that contribute to perturbation series. The analytical expressions for corresponding diagrams read:
\begin{eqnarray}
&&Z_1=\frac{u(2\pi)^{-d/2}L^{2-d/2}}{(1-d/2)(2-d/2)},\\
&&Z_2=\frac{u(2\pi)^{-d/2}L^{2-d/2}(2^{2-d/2}-2)}{(1-d/2)(2-d/2)},\\
&&Z_3=\frac{u(2\pi)^{-d/2}L^{2-d/2}(3^{2-d/2}-2\,2^{2-d/2}+1)}{(1-d/2)(2-d/2)}.\\
&&Z_4=\frac{u(2\pi)^{-d/2}L^{2-d/2}(4^{2-d/2}-2\,3^{2-d/2}+2^{2-d/2})}{(1-d/2)(2-d/2)}.
\end{eqnarray}
All of these expressions are considered as series expansions based on the deviation $\epsilon=4-d$ from the upper critical dimension of the coupling constant $u$. In this context, the contributions can be expressed as follows:
\begin{eqnarray}
&&Z_1=-\frac{2}{\epsilon}-1,\\
&&Z_2=-\frac{2}{\epsilon}-1-\ln(2),\\
&&Z_3=2\ln(2)-\ln(3),\\
&&Z_4=-3\ln(2)+2\ln(3).
\end{eqnarray}
Note that each of the diagrams needs to be taken into account with the corresponding combinatorial pre-factor. The resulting expression for a  partition function reads:
\begin{eqnarray}
&&Z=1-u_0\left(\frac{1}{\epsilon}\left(f_s(f_s-1)+\sum_{i=1}^{f_s} (f_i^2-3f_i)\right)\right.\nonumber\\
&&+\frac{1}{2}\left(f_s(f_s-1)+\sum_{i=1}^{f_s} (f_i^2-3f_i)+\ln(2)\left(f_s(1-f_s)\right.\right.\nonumber\\
&&\left.\sum_{i=1}^{f_s}\left(4(f_s-1)-f_i-3\sum_{j=1}^{f_s}(f_j-1)\right)(f_i-1)\right)\\
&&\left.\left.+\ln(3)\sum_{i=1}^{f_s}\left((f_i-1)\left(1-f_s+\sum_{j=1}^{f_s}(f_j-1)\right)\right)\right)\right)\label{Z1loop}\nonumber
\end{eqnarray}
where $u_0=u(2\pi)^{-d/2}L^{2-d/2}$ is a dimensionless coupling constant.
Subsequently, we assume that all functionalities of external branching have equal values, denoted as $f_i=f$ for $i=1,\ldots,f_s$.

\subsection{Radius of gyration}

\begin{figure}
\begin{center}
\includegraphics[width=73mm]{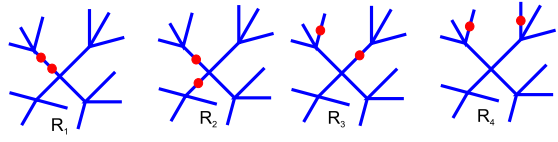}
\caption{ \label{fig:3}Diagrammatic representation of contributions into the radius of gyration in Gaussian approximation.  The solid lines depict the paths of the polymers, each with a length of $L$, while the bullets symbolize the restriction points. }
\end{center}
\end{figure}

The mean-square radius of gyration of a snowflake-shaped macromolecule in the continuous chain model  is defined by \begin{eqnarray}
&&\langle {R^2_{g}}\rangle_H = \frac{1}{2L^2(\sum_{i=1}^{f_s} f_i)^2} \times \nonumber \\
&&\times \sum_{i,j=1}^{F}\int_0^L\int_0^L ds_1\,ds_2 \langle(\vec{r}_i(s_2)-\vec{r}_j(s_1))^2\rangle.
\end{eqnarray}
Here and below, $\langle \ldots \rangle_H$ denotes averaging with the Hamiltonian of Eq.~(\ref{H}) according to:
\begin{eqnarray}
&&\langle (\ldots) \rangle = \frac{1}{ Z^{\rm snowflake}_{\{f_i\},f_s}}\prod_{i=1}^{f_s}\prod_{j=2}^{f_i}\,\int\,D\vec{r}(s)\,\times\nonumber\\
&&\times\delta(\vec{r^i_1}(L))\delta(\vec{r^i_j}(0)-\vec{r^i_1}(0))\,{\rm e}^{-H}(\ldots).
\end{eqnarray}
The mean-square distance between any two points along the trajectories (so-called restriction points $s_1$ and $s_2$) are calculated according to the formula:
\begin{eqnarray}
&&\langle(\vec{r}_i(s_2)-\vec{r}_j(s_1))^2\rangle_H = - 2 \frac{d}{d|\vec{k}|^2}\xi(\vec{k})_{\vec{k}=0},\nonumber\\
&&\xi(\vec{k})\equiv\langle{\rm e}^{-\iota\vec{k}(\vec{r}_i(s_2)-\vec{r}_j(s_1))}\rangle_H.
\end{eqnarray}
We employ a path integration approach to evaluate $\xi(\vec{k})$ using perturbation theory expansions, similar to how it was performed for the partition function in the preceding subsection.   Once more we use a diagrammatic presentation which we display in Fig.~\ref{fig:3}. In the case of Gaussian approximation, corresponding analytical expressions are given by:
\begin{eqnarray}
R_1=\frac{1}{6},\, R_2=1, \,R_3=2, \,R_4=3.
\end{eqnarray}
Taking into account the combinatorial pre-factors of diagrams  we get the analytical expression for the radius of in the Gaussian approximation:
\begin{eqnarray}
&&\langle R^2_g\rangle_0 =\frac{dL}{6(f_s f)} (3f_s-2)(3f-2). \label{rgGauss}
\end{eqnarray}

\begin{figure}
\begin{center}
\includegraphics[width=53mm]{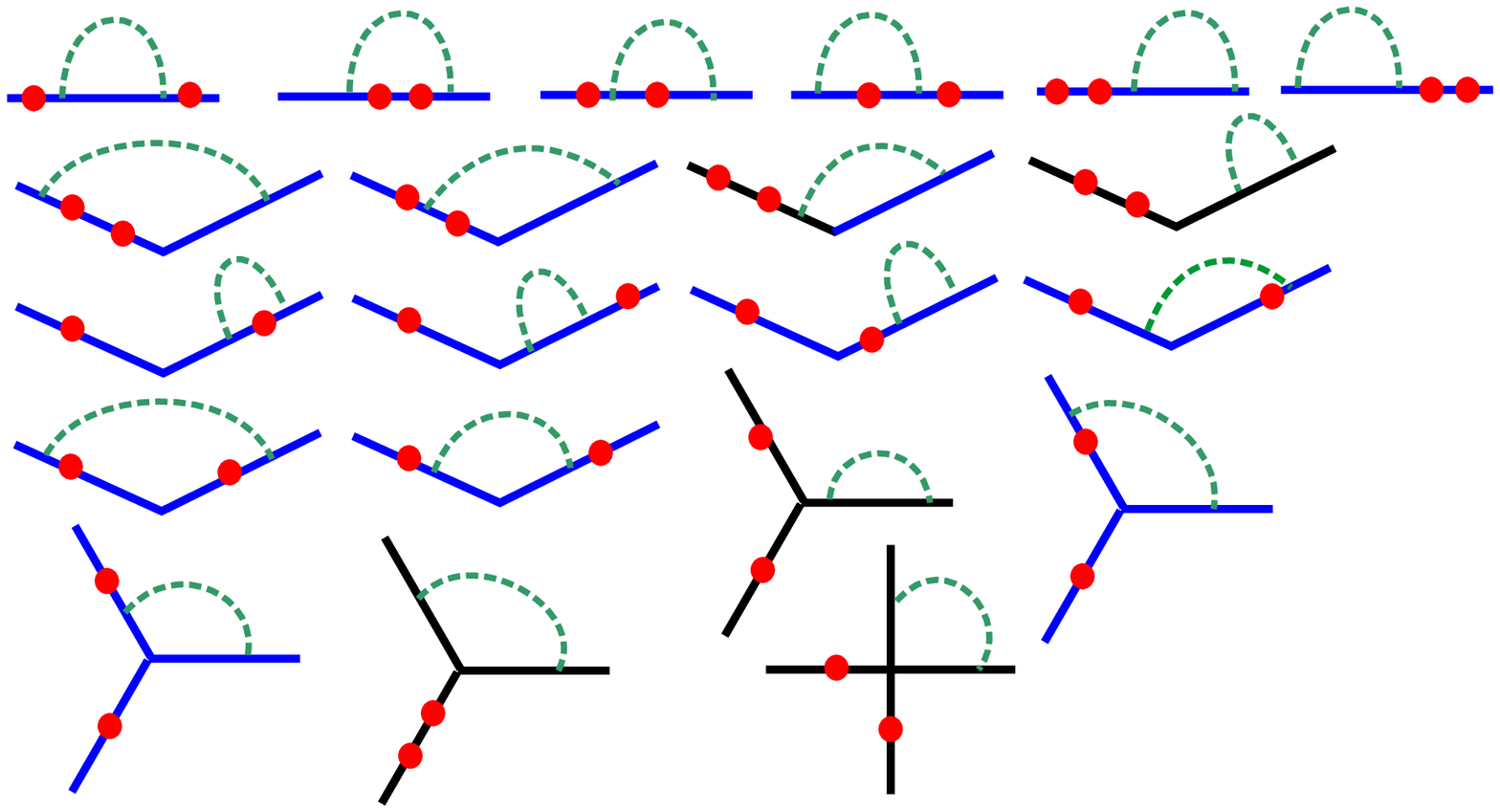}
\includegraphics[width=53mm]{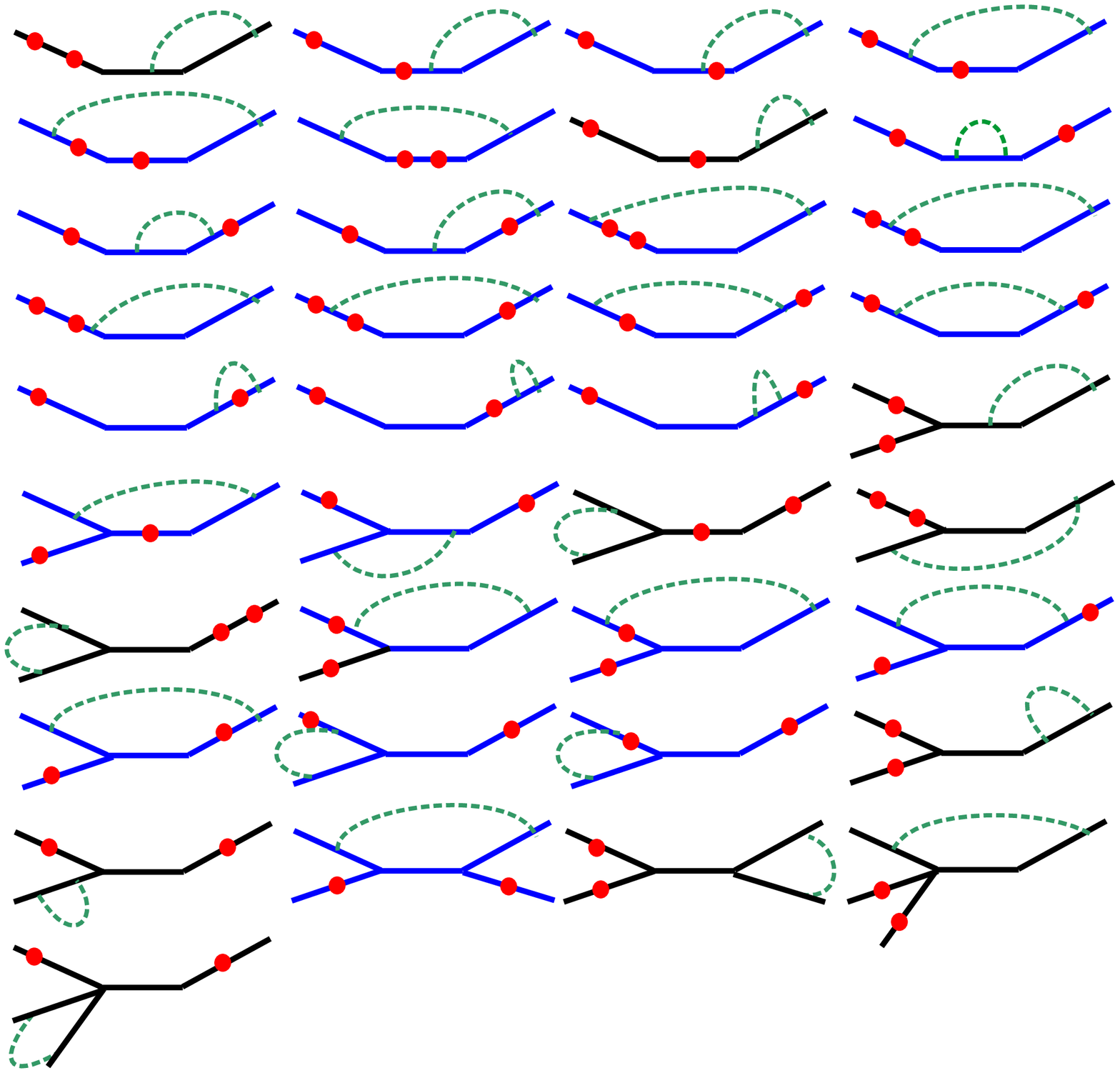}
\includegraphics[width=53mm]{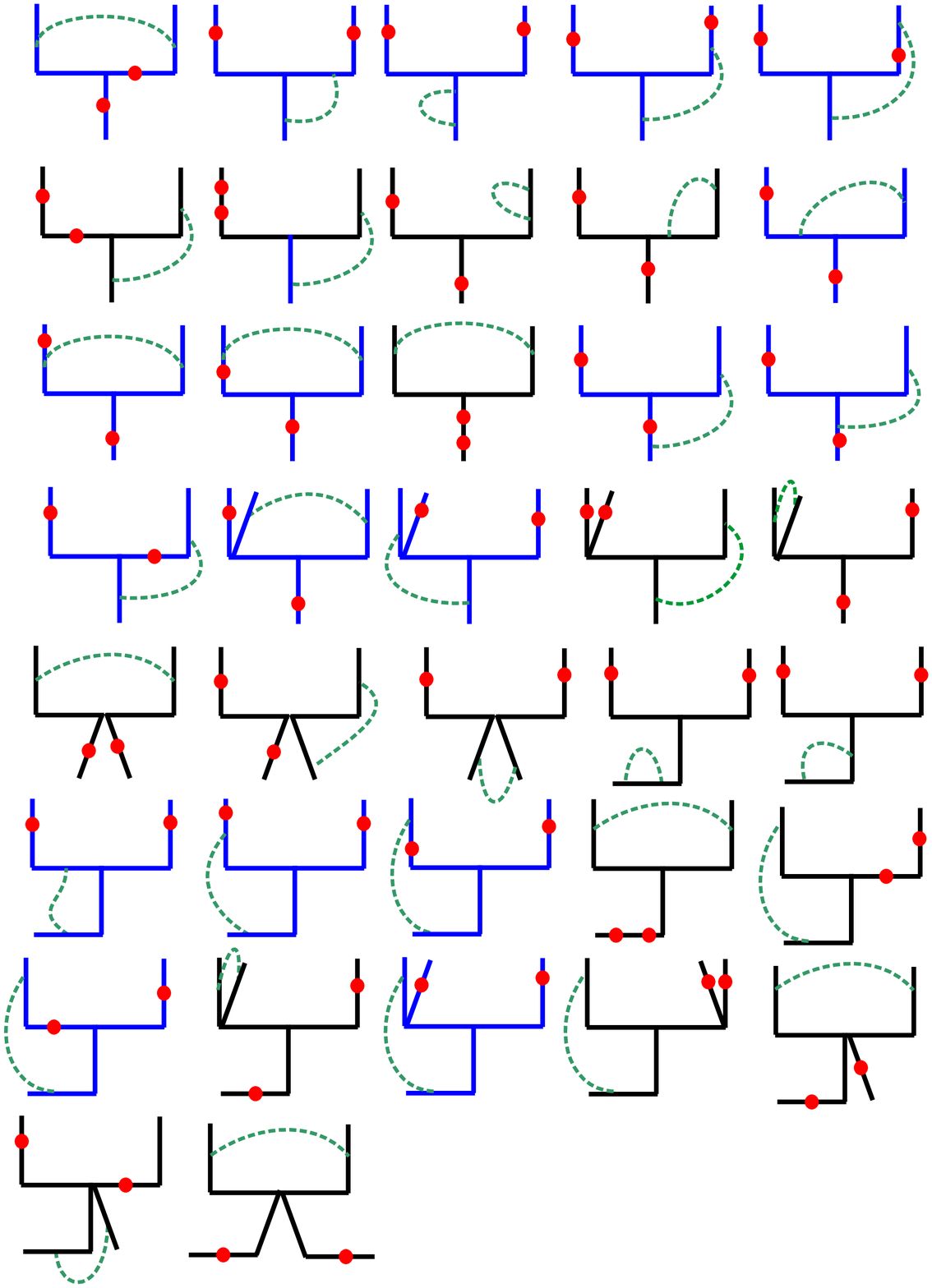}
\caption{ \label{Snow_D_1l} Diagrammatic presentation of the contributions into the gyration radius in one loop approximation, i.e. with included excluded volume interactions. The solid lines denote polymer trajectories, red bullets denote restriction points and the dash lines represent a two point excluded volume interactions.}
\end{center}
\end{figure}

To evaluate the expression for the radius of gyration with taking into account the excluded volume interactions, 
we again make use of diagrammatic technique. By summing up the contributions of all diagrams depicted in Fig.~\ref{Snow_D_1l}, we obtain the expression for $d=3$:
  \begin{eqnarray}
  &&\langle R^2_g \rangle =-\frac{1}{105f(3g-2)(3f-2)}\left(27840+12400f-34720\sqrt{3}gf^4-121661gf-149340f^2-23310g^2+314217f^2g\right.\nonumber\\
  &&-153419f^2g^2-4530g-11235f^4g^2+34545f^4g+23520\sqrt{3}f^4-11970\sqrt{2}f^4-110880\sqrt{3}f^3+45570\sqrt{2}f^3\nonumber\\
  &&+150170\sqrt{3}f^2-75530\sqrt{2}f^2-27150\sqrt{3}f+22350\sqrt{2}f-23310f^4+57183\sqrt{2}g^2f-74413\sqrt{2}gf+124950f^3\nonumber\\
  &&+24700\sqrt{2}+180096\sqrt{3}gf^3-68495\sqrt{2}gf^3-69216\sqrt{3}g^2f^3+133662\sqrt{3}gf-291178\sqrt{3}gf^2+25753\sqrt{2}g^2f^3\nonumber\\
  &&+130155\sqrt{2}gf^2+11200\sqrt{3}g^2f^4+141008\sqrt{3}g^2f^2-62573\sqrt{2}g^2f^2+17535\sqrt{2}gf^4-106512\sqrt{3}g^2f-5565\sqrt{2}g^2f^4\nonumber\\
  &&+12140\sqrt{3}g+23520\sqrt{3}g^2-12730\sqrt{2}g-11970\sqrt{2}g^2+82299f^3g\left.^2-211113f^3g-35660\sqrt{3}+101801fg^2\right)
 \end{eqnarray}

\begin{figure}[b!]
\begin{center}
\includegraphics[width=85mm]{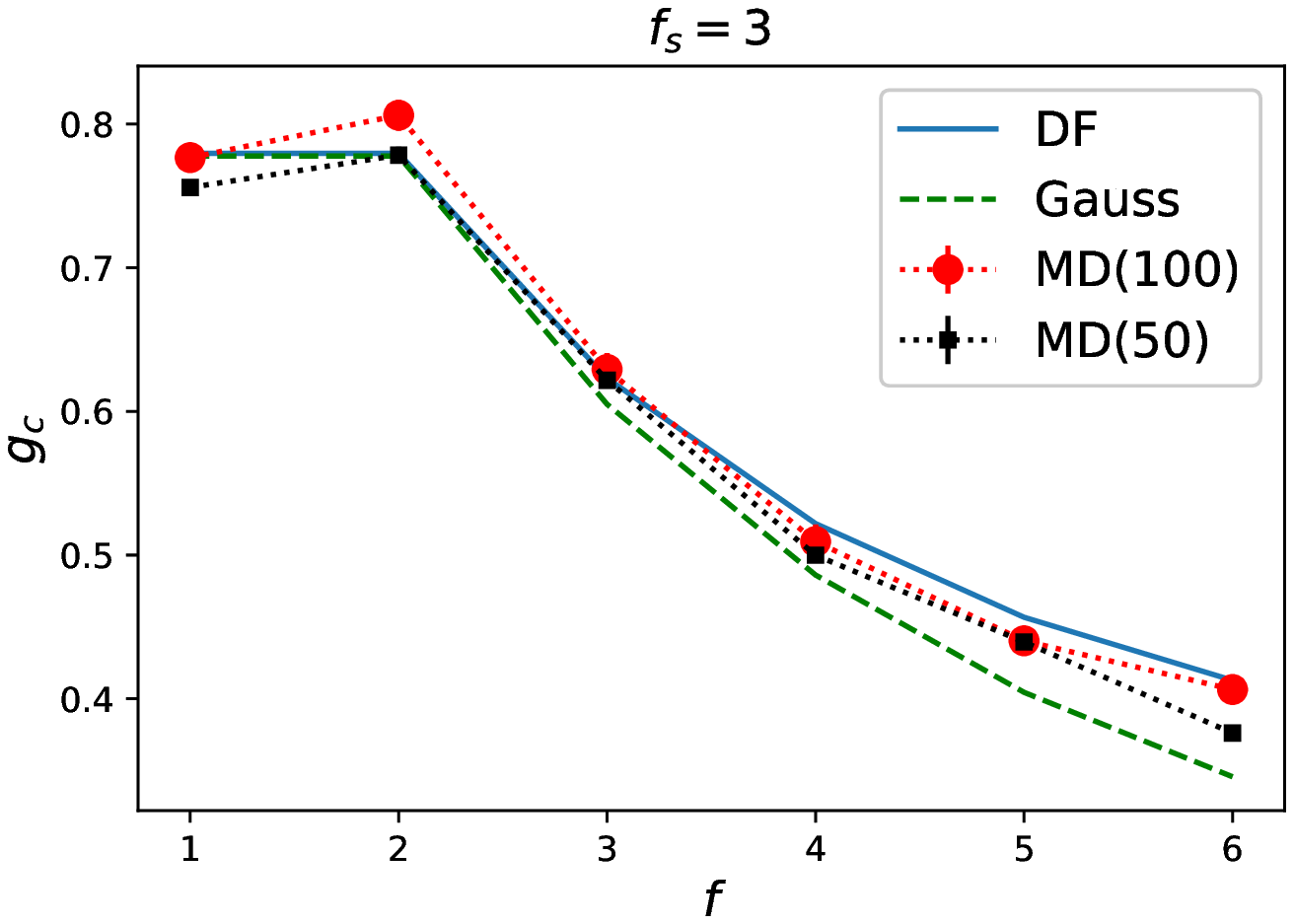}
\includegraphics[width=85mm]{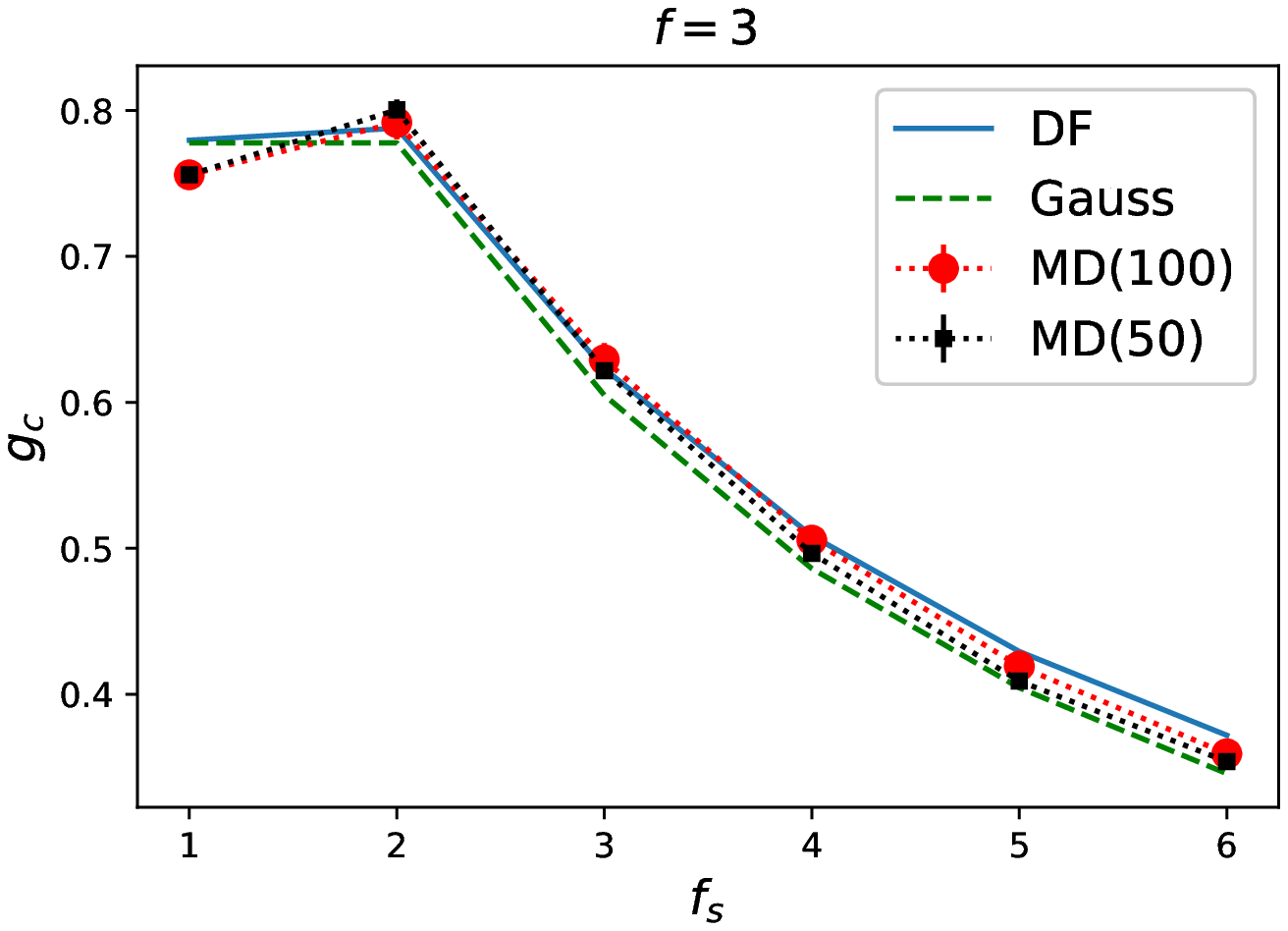}
\caption{ \label{sizeratio}A size ratio $g_c$ as a function of branching parameters: $f$ for fixed $f_s=3$ (on the top) and $f_s$ for the fixed $f=3$ (on the bottom) at $d=3$.}
\end{center}
\end{figure}

\subsection{Size ratios}

To evaluate the size ratio $g_{\rm complex}$ of  Eq.~(\ref{gratio}) for a snowflake-shaped polymer in 
 Gaussian approximation, we  use of the expression (\ref{rgGauss}) and recall that the radius of gyration of a linear chain of the same total molecular weight $f_sfL$ is: ${\langle R_g^2 \rangle_{\rm chain}}_0=dLf_sf/6$. The final expression for size ratio reads:
\begin{eqnarray}
g_c=\frac{(3f_s-2)(3f-2)}{(f_sf)^2}. \label{ratioGauss}
\end{eqnarray}
By utilizing this expression, we can recover the established values for the single-branched star polymer (\ref{gstar}) when $f=1$. Additionally, the results obtained for dendrimers in the first generation, as reported in references \cite{Boris96, Ferla97}, can be reproduced.   

To incorporate excluded volume interactions, we employ the Douglas-Freed approximation \cite{Douglas84}. In this approximation, a size ratio can be defined as follows: 
\begin{equation}
g_c=\frac{\langle R^2_g\rangle_0}{\langle R^2_g\rangle_{0,\rm linear}}\frac{1-a_{\rm snowflake}}{1-a_{\rm linear}}    
\end{equation}
with $a_{x}=\frac{3C}{32}-\frac{1}{4}$ where $C$ is a coefficient in $\langle R^2_g\rangle=\langle R^2_g\rangle_0(1-u C)$ for $d=3$. The final expression for a snowflake polymer with  $f_i=f$ reads:
\begin{eqnarray}
&&g_c=\frac{420}{487}\frac{(3f_s-2)(3f-2)}{(f^2f_s^2)}-\frac{1}{974}\frac{1}{f^3\sqrt{f_sf}f_s^2}\times.\nonumber\\
&&\left(-27840+62573\sqrt{2}f^2f_s^2-141008\sqrt{3}f^2f_s^2\right.\nonumber\\ &&\left.+291178\sqrt{3}f^2f_s+74413\sqrt{2}ff_s-25753\sqrt{2}f^3f_s^2\right.\nonumber\\ &&\left.+34720\sqrt{3}f^4f_s+68495\sqrt{2}f^3f_s+5565\sqrt{2}f^4f_s^2\right.\nonumber\\ &&\left.-12400f+69216\sqrt{3}f^3f_s^2-57183\sqrt{2}ff_s^2\right.\nonumber\\ &&\left.-130155\sqrt{2}f^2f_s-11200\sqrt{3}f^4f_s^2+106512\sqrt{3}ff_s^2\right.\nonumber\\ &&\left.-133662\sqrt{3}ff_s-180096\sqrt{3}f^3f_s-17535\sqrt{2}f^4f_s\right.\nonumber\\ &&\left.+153419f^2f_s^2-150170\sqrt{3}f^2+75530\sqrt{2}f^2\right.\nonumber\\ &&\left.-45570\sqrt{2}f^3-22350\sqrt{2}f-23520\sqrt{3}f^4+23310f_s^2\right.\nonumber\\ &&\left.+110880\sqrt{3}f^3+11970\sqrt{2}f^4+27150\sqrt{3}f+4530f_s\right.\nonumber\\ &&\left.+149340f^2-124950f^3+23310f^4+211113f^3f_s\right.\nonumber\\ &&\left.+11235f^4f_s^2-82299f^3f_s^2-314217f^2f_s-101801ff_s^2\right.\nonumber\\ &&\left.+121661f_sf+11970\sqrt{2}f_s^2-23520\sqrt{3}f_s^2+12730\sqrt{2}f_s\right.\nonumber\\ &&\left.-12140\sqrt{3}f_s-34545f^4f_s-24700\sqrt{2}+35660\sqrt{3}\right)
\end{eqnarray}
Note that putting $f=1$ in the above expression, we recover the size ratio of a star polymer with excluded volume interactions \cite{Blavatska12}. When $f_s=2$, we can restore the pom-pom structure, where the length of the central backbone is twice as large as the lengths of the side chains.
The obtained quantitative results based on this expression for a fixed $d=3$, along with the numerical values obtained by our calculations, are presented in Tables \ref{tab:1} and \ref{tab:2}. These results are also depicted graphically in Fig.~\ref{sizeratio} for a specific set of parameter values $f_s$ and $f$. We observe that our numerical data show reasonably good agreement with the analytical predictions, with the best agreement observed for the structure with $f_s=f=3$. As the branching increases, the finite size effects in numerical simulations become more noticeable.  
The main conclusion that can be drawn is that as the parameter $f_s$ increases, the value of $g_c$ decreases. This indicates that the effective size of the snowflake-shaped polymer  becomes smaller when compared to a linear polymer coil of the same total molecular weight. In other words, increasing the branching of the central core leads to more compact conformations of snowflake polymers.

\begin{table}[b!]
    \begin{tabular}{|c|c|c|c|c|}
    \hline
        $f$ & $3$ & $4$ & $5$ & $6$ \\ \hline
        DF & $0.623$&$0.5219$&$0.4567$&$0.4128$\\ \hline	
        MD(100)& $0.629(11)$& $0.509(12)$&$0.440(9)$&$0.406(9)$\\ \hline
        MD(50)& $0.0.622(4)$& $0.4999(35)$&$0.439(3)$&$0.376(3)$\\ \hline
        Gauss, RG &$0.60494$&	$0.48611$&$0.40444$&$0.34568$\\ \hline
        Gauss, Wei's method &$0.606(1) $&	$0.489(1) $ &$0.403(1) $&$0.346(1) $\\ \hline
            \end{tabular}
    \caption{The values for the size ratio $g_c$ for $f_s=3$}
    \label{tab:1}
\end{table}

\begin{table}[b!]

    \begin{tabular}{|c|c|c|c|c|}
    \hline
        $f_s$ & $3$ & $4$ & $5$ & $6$ \\ \hline
        DF & $0.623$&$0.5086$&$0.4293$&$0.3719$\\ \hline	
        MD(100)& $0.629(11)$& $0.507(9)$&$0.419(6)$&$0.359(6)$\\ \hline
        MD(50)& $0.0.622(4)$& $0.497(3)$&$0.409(2)$&$0.354(1)$\\ \hline
        Gauss, RG &$0.60494$&	$0.48611$&$0.40444$&$0.34568$\\ \hline
        Gauss, Wei's method &$ 0.606(1)$&	$ 0.487(1) $
        &$ 0.405(1)$& $ 0.346(1)  $\\ \hline
            \end{tabular}
    \caption{The values for the size ratio $g_c$ for $f=3$}
    \label{tab:2}
\end{table}

\subsection{Asphericity}

\begin{figure}[t!]
\begin{center}
\includegraphics[width=85mm]{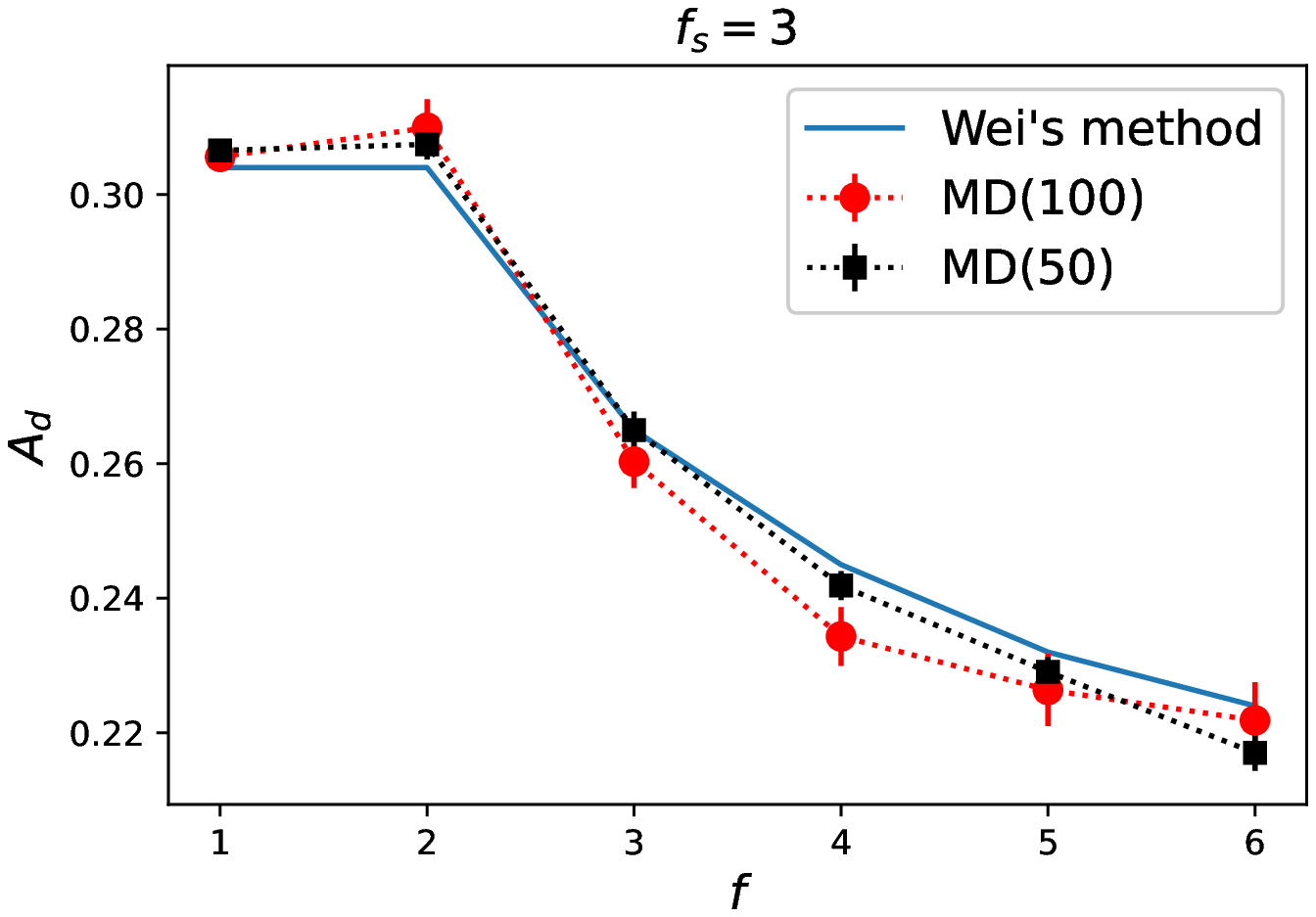}
\includegraphics[width=85mm]{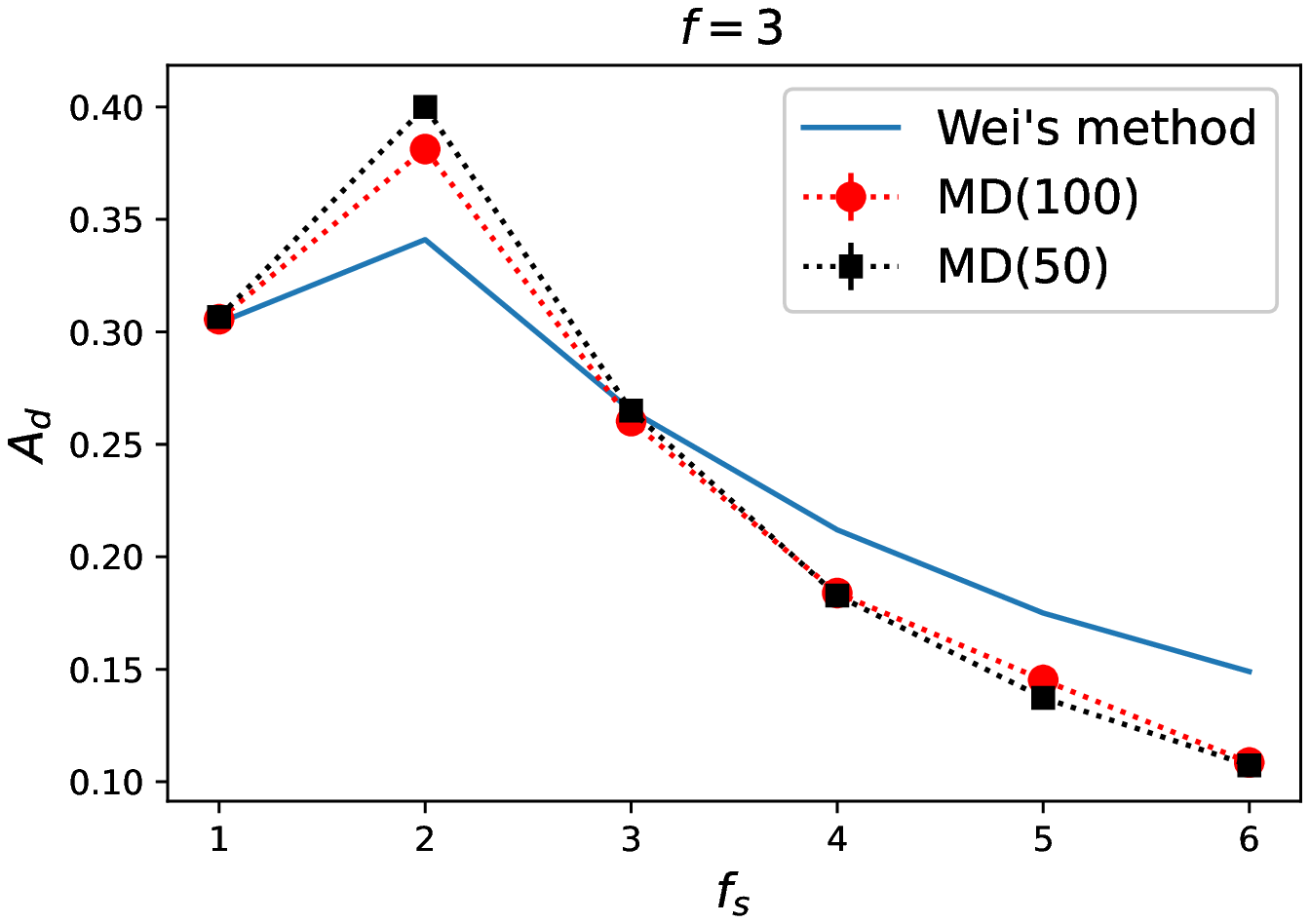}
\caption{ \label{Af} The averaged asphericity $A_3$ as a function of branching parameters: $f$ for fixed $f_s=3$ (on the top) and $f_s$ for the fixed $f=3$ (on the bottom).}
\end{center}
\end{figure}

Due to the complexity and difficulty in handling the highly intricate expressions arising from analytical calculations of asphericity (as defined by Eq. (\ref{Ad})), our focus was primarily on numerically evaluating this quantity.
The results for averaged asphericity $\langle A_d \rangle$ obtained by molecular dynamics and Wei's method are presented graphically in Fig. \ref{Af}.  Let's start by examining the case where the functionality of the central branching point is fixed at $f_s=3$. For $f=1$ and $f=2$, we can determine the asphericity of a star polymer with three branches using Wei's method. As the value of $f$ increases, the asphericity parameter decreases, indicating an increase in the symmetry of the snowflake-like architecture.
Now, let's analyze the results for a fixed value of $f_s=3$. When $f=1$, we restore the case of a simple star with three arms. For $f_s=2$, we obtain the pom-pom polymer structure, which is known to be more asymmetric compared to a star polymer with the same total molecular weight (as indicated by the higher asphericity parameter). However, with further increases in $f_s$, the asphericity decreases, suggesting that the presence of side branching makes the polymer structure more symmetric than a conventional star polymer.
It is important to note that the impact of excluded volume interactions is relatively small for cases with low inner branching ($f_s=3$). However, as $f_s$ increases, the excluded volume interactions become more significant, resulting in a more spherical shape than the prediction from the Gaussian approximation obtained using Wei's method.
\section{Conclusions}
\label{Con}

In this study, we have conducted an analytical and numerical investigation of a multibranched polymer structure known as a snowflake polymer. This polymer architecture consists of a central core with functionality $f_s$ and dendronized side arms with functionalities $f$. Recently, a similar structure has been synthesized and characterized as a promising alternative to complex high-generation dendrimers with comparable functional properties \cite{Liu21}.

Our focus was on exploring the universal size and shape characteristics of this polymer topology in a dilute, good solvent regime. Specifically, we examined the size ratio (defined by Eq. (\ref{gratio})) and asphericity (given by Eq. (\ref{Ad})) of typical polymer conformations. Our results reveal a quantitative decrease in the size ratio and asphericity as the branching parameters increase, indicating a compactification effect on the complex polymer structure compared to a linear coil with the same total molecular weight.

Furthermore, we observed that an increase in the inner branching parameter ($f_s$) leads to a more compact and spherical polymer structure compared to when the number of side branchings ($f$) is increased. For example, when $f_s=3$ and $f=6$, the size ratio is approximately 0.41, while for the reverse case of $f=3$ and $f_s=6$, the size ratio is approximately 0.36. The difference in asphericity is even more pronounced in these two cases.

Overall, our findings provide valuable insights into the structural properties of snowflake polymers and demonstrate the impact of branching parameters on their compactness and shape.

\begin{acknowledgements}
K.H. and J.P. would like to acknowledge the support from the National Science Center, Poland (Grant No. 2018/30E/ST3/00428) and the computational time at PL-Grid,  Poland.
\end{acknowledgements}

\bibliographystyle{unsrt}
\bibliography{Pom-Pom}

\begin{thebibliography}{10}

\bibitem{Duro2015}
A.~Duro-Castano, J.~Movellan, and M.~J. Vicent.
\newblock Smart branched polymer drug conjugates as nano-sized drug delivery
  systems.
\newblock {\em Biomater. Sci.}, 3:1321--1334, 2015.

\bibitem{England2010}
Richard~Mark England and Stephen Rimmer.
\newblock Hyper/highly-branched polymers by radical polymerisations.
\newblock {\em Polym. Chem.}, 1:1533--1544, 2010.

\bibitem{Higashihara11}
Tomoya Higashihara, Yukari Segawa, Warapon Sinananwanich, and Mitsuru Ueda.
\newblock Synthesis of hyperbranched polymers with controlled degree of
  branching.
\newblock {\em Polymer Journal}, pages 155--160, 2012.

\bibitem{Zheng15}
Yaochen Zheng, Sipei Li, Zhulin Weng, and Chao Gao.
\newblock Hyperbranched polymers: advances from synthesis to applications.
\newblock {\em Chem. Soc. Rev.}, 44:4091--4130, 2015.

\bibitem{polymeropoulos2017}
George Polymeropoulos, George Zapsas, Konstantinos Ntetsikas, Panayiotis
  Bilalis, Yves Gnanou, and Nikos Hadjichristidis.
\newblock 50th anniversary perspective: Polymers with complex architectures.
\newblock {\em Macromolecules}, 50(4):1253--1290, 2017.

\bibitem{McLeish98}
T.~C.~B. McLeish and R.~G. Larson.
\newblock Molecular constitutive equations for a class of branched polymers:
  The pom-pom polymer.
\newblock {\em Journal of Rheology}, 42(1):81--110, January 1998.

\bibitem{Saying14}
Wuliji Saiyin, Dali Wang, Lili Li, Lijuan Zhu, Bing Liu, Lijian Sheng, Yanwu
  Li, Bangshang Zhu, Limin Mao, Guolin Li, and Xinyuan Zhu.
\newblock Sequential release of autophagy inhibitor and chemotherapeutic drug
  with polymeric delivery system for oral squamous cell carcinoma therapy.
\newblock {\em Molecular Pharmaceutics}, 11(5):1662--1675, 2014.
\newblock PMID: 24666011.

\bibitem{Yu14}
Songrui Yu, Ruijiao Dong, Jianxin Chen, Feng Chen, Wenfeng Jiang, Yongfeng
  Zhou, Xinyuan Zhu, and Deyue Yan.
\newblock Synthesis and self-assembly of amphiphilic aptamer-functionalized
  hyperbranched multiarm copolymers for targeted cancer imaging.
\newblock {\em Biomacromolecules}, 15(5):1828--1836, 2014.
\newblock PMID: 24750012.

\bibitem{Carter06}
Steven Carter, Stephen Rimmer, Ramune Rutkaite, Linda Swanson, J.~P.~A.
  Fairclough, Alice Sturdy, and Michelle Webb.
\newblock Highly branched poly(n-isopropylacrylamide) for use in protein
  purification.
\newblock {\em Biomacromolecules}, 7(4):1124--1130, 2006.
\newblock PMID: 16602729.

\bibitem{Zimm49}
Bruno~H. Zimm and Walter~H. Stockmayer.
\newblock The dimensions of chain molecules containing branches and rings.
\newblock {\em The Journal of Chemical Physics}, 17(12):1301--1314, dec 1949.

\bibitem{Graham01}
R.~S. Graham, T.~C.~B. McLeish, and O.~G. Harlen.
\newblock Using the pom-pom equations to analyze polymer melts in exponential
  shear.
\newblock {\em Journal of Rheology}, 45(1):275--290, January 2001.

\bibitem{Ruymbeke07}
Evelyne van Ruymbeke, Michael Kapnistos, Dimitris Vlassopoulos, Tianzi Huang,
  and Daniel~M. Knauss.
\newblock Linear melt rheology of pom-pom polystyrenes with unentangled
  branches.
\newblock {\em Macromolecules}, 40(5):1713--1719, March 2007.

\bibitem{Chen11}
Xue Chen, M.~Shahinur Rahman, Hyojoon Lee, Jimmy Mays, Taihyun Chang, and
  Ronald Larson.
\newblock Combined synthesis, {TGIC} characterization, and rheological
  measurement and prediction of symmetric h polybutadienes and their blends
  with linear and star-shaped polybutadienes.
\newblock {\em Macromolecules}, 44(19):7799--7809, October 2011.

\bibitem{1996}
Wolfgang Radke and Axel H.~E. M\"{u}ller.
\newblock Mean square radius of gyration and hydrodynamic radius of jointed
  star (dumbbell) and h-comb polymers.
\newblock {\em Macromolecular Theory and Simulations}, 5(4):759--769, July
  1996.

\bibitem{Kalyuzhnyi20}
K.~Haydukivska, O.~Kalyuzhnyi, V.~Blavatska, and Ja. Ilnytskyi.
\newblock On the swelling properties of pom-pom polymers in dilute solutions.
  part 1: Symmetric case.
\newblock {\em Journal of Molecular Liquids}, 328:115456, April 2021.

\bibitem{Haydukivska22}
K.~Haydukivska, O.~Kalyuzhnyi, V.~Blavatska, and Ja. Ilnytskyi.
\newblock Swelling of asymmetric pom-pom polymers in dilute solutions.
\newblock {\em Condensed Matter Physics,}, 25(2):23302, 2022.

\bibitem{Sheiko}
S.~Sheiko, B.S. Sumerlin, and K.~Matyjaszewski.
\newblock Cylindrical molecular brushes: Synthesis, characterization, and
  properties.
\newblock {\em Progress in Polymer Science}, 33(7):759--785, 2008.

\bibitem{paturej1}
Jaros{\l}aw Paturej and Torsten Kreer.
\newblock Hierarchical excluded volume screening in solutions of bottlebrush
  polymers.
\newblock {\em Soft matter}, 13(45):8534--8541, 2017.

\bibitem{paturej2}
Jaros{\l}aw Paturej, Sergei~S Sheiko, Sergey Panyukov, and Michael Rubinstein.
\newblock Molecular structure of bottlebrush polymers in melts.
\newblock {\em Science advances}, 2(11):e1601478, 2016.

\bibitem{paturej3}
William~FM Daniel, Joanna Burdy{\'n}ska, Mohammad Vatankhah-Varnoosfaderani,
  Krzysztof Matyjaszewski, Jaros{\l}aw Paturej, Michael Rubinstein, Andrey~V
  Dobrynin, and Sergei~S Sheiko.
\newblock Solvent-free, supersoft and superelastic bottlebrush melts and
  networks.
\newblock {\em Nature materials}, 15(2):183--189, 2016.

\bibitem{Terao}
K.~Terao, T.~Hokajo, Y.~Nakamura, and T.~Norisuye.
\newblock Solution properties of polymacromonomers consisting of polystyrene.
  3. viscosity behavior in cyclohexane and toluene.
\newblock {\em Macromolecules}, 32(11):3690--3694, 1999.

\bibitem{Kawaguchi}
S.~Kawaguchi, K.~Akaike, Z.-M. Zhang, H.~Matsumoto, and K.~Ito.
\newblock Water soluble bottlebrushes.
\newblock {\em Polym. J.}, 30:1004, 1998.

\bibitem{Vogtle}
E.~Buhleier, W.~Wehner, and F.~Vogtle.
\newblock Cascade and nonskid-chain-like synthesis of molecular cavity
  topologies.
\newblock {\em Synthesis}, 2:155--158, 1978.

\bibitem{Tomalia}
D.A. Tomalia and et~al.
\newblock A new class of polymers: starburst-dendritic macromolecules.
\newblock {\em Polym J}, 17:117--132, 1985.

\bibitem{Mathur10}
Vineet Mathur, Yamini Satrawala, and Mithun Rajput.
\newblock Dendrimers: A review.
\newblock {\em Inventi Impact: NDDS}, 1:14, 01 2010.

\bibitem{Liu21}
Y.~Liu, S.~Bai, T.~Wu, C.C. Chen, Y.~Liu, X.~Chao, and Y.~Bai.
\newblock Dendronized arm snowflake polymer as a highly branched scaffold for
  cellular imaging and delivery.
\newblock {\em Biomacromolecules}, 22:3791–3799, 2021.

\bibitem{wang}
Zhen-Gang Wang.
\newblock 50th anniversary perspective: Polymer conformation: A pedagogical
  review.
\newblock {\em Macromolecules}, 50(23):9073--9114, 2017.

\bibitem{burchard}
Walther Burchard.
\newblock {\em Solution Properties of Branched Macromolecules}, pages 113--194.
\newblock Springer Berlin Heidelberg, Berlin, Heidelberg, 1999.

\bibitem{desCloiseaux}
Jaques des Cloizeaux and Gerard Jannink.
\newblock {\em Polymers in Solution: their modelling and structure}.
\newblock Clarendon Press:Oxford, 1991.

\bibitem{Douglas84}
Jack Douglas and Karl~F. Freed.
\newblock Renormalization and the two-parameter theory.
\newblock {\em Macromolecules}, 17(11):2344--2354, November 1984.

\bibitem{Paturej22}
Khristine Haydukivska, Viktoria Blavatska, Jaros\l{}aw~S. K\l{}os, and
  Jaros\l{}aw Paturej.
\newblock Conformational properties of hybrid star-shaped polymers comprised of
  linear and ring arms.
\newblock {\em Phys. Rev. E}, 105:034502, Mar 2022.

\bibitem{Kaluzhniy22}
K.~Haydukivska, O.~Kalyuzhnyi, V.~Blavatska, and Ja. Ilnytskyi.
\newblock Swelling of asymmetric pom-pom polymers in dilute solutions.
\newblock {\em Condensed Matter Physics}, 25:23302, 2022.

\bibitem{Radke96}
W.~Radke and M\"uller A.~H. E.
\newblock Mean square radius of gyration and hydrodynamic radius of jointed
  star (dumbbell) and h-comb polymers.
\newblock {\em Macromolecular Theory and Simulations}, 5:759--769, 1996.

\bibitem{Nakamura}
Yo~Nakamura, Yunan Wan, Jimmy~W. Mays, Hermis Iatrou, and Nikos
  Hadjichristidis.
\newblock Radius of gyration of polystyrene combs and centipedes in solution.
\newblock {\em Macromolecules}, 33(22):8323--8328, 2000.

\bibitem{Boris96}
D.~Boris and M.~Rubinstein.
\newblock A self-consistent mean field model of a starburst dendrimer: Dense
  core vs dense shell.
\newblock {\em Macromolecules}, 29:7251--7260, 1996.

\bibitem{Ferla97}
R.~La~Ferla.
\newblock Conformations and dynamics of dendrimers and cascade macromolecules.
\newblock {\em J. Chem. Phys.}, 106:688--700, 1997.

\bibitem{Timoshenko02}
Edward~G. Timoshenko, Yuri~A. Kuznetsov, and Ronan Connolly.
\newblock Conformations of dendrimers in dilute solution.
\newblock {\em The Journal of Chemical Physics}, 117(19):9050--9062, 2002.

\bibitem{Sheng08}
Yu-S. Sheng, S.~Jiang, and H.K. Tsao.
\newblock Radial size of a starburst dendrimer in solvents of varying quality.
\newblock {\em Macromolecules}, 35:865--7868, 2002.

\bibitem{Ganazzoli00}
F.~Ganazzoli, R.~La~Ferla, and G.~Terragni.
\newblock Conformational properties and intrinsic viscosity of dendrimers under
  excluded-volume conditions.
\newblock {\em Macromolecules}, 33:6611–6620, 2000.

\bibitem{Ganazzoli02}
F.~Ganazzoli.
\newblock Conformations and dynamics of stars and dendrimers: the gaussian
  self-consistent approach.
\newblock {\em Condensed Matter Phys.}, 5:37--71, 2002.

\bibitem{Mansfield00}
M.L. Mansfield.
\newblock Monte carlo studies of dendrimers. additional results for the diamond
  lattice model.
\newblock {\em Macromolecules}, 33:8043–8049, 2000.

\bibitem{Tande01}
B.M. Tande, M.J. Wagner, M.E. Mackay, C.J. Hawker, and M.~Jeong.
\newblock Viscosimetric, hydrodynamic, and conformational properties of
  dendrimers and dendrons.
\newblock {\em Macromolecules}, 34:8580–8585, 2001.

\bibitem{Aronovitz}
J.A. Aronovitz and D.R. Nelson.
\newblock Universal features of polymer shapes.
\newblock {\em J. Physique}, 47:1445--1456, 1986.

\bibitem{Rudnick86}
J.~Rudnick and G.~Gaspari.
\newblock The aspherity of random walks.
\newblock {\em J. Phys. A}, 19:L191--L194, 1986.

\bibitem{Blavatska11}
V.~Blavatska, C.~von Ferber, and Yu. Holovatch.
\newblock Shapes of macromolecules in good solvents: field theoretical
  renormalization group approach.
\newblock {\em Condens. Matter Phys.}, 14:33701: 1--20, 2011.

\bibitem{Ferber15}
Christian von Ferber, Marvin Bishop, Thomas Forzaglia, Cooper Reid, and Gregory
  Zajac.
\newblock The shapes of simple three and four junction comb polymers.
\newblock {\em The Journal of Chemical Physics}, 142(2):024901, January 2015.

\bibitem{Blavatska20}
V.~Blavatska, K.~Haydukivska, and Yu. Holovatch.
\newblock Shape analysis of random polymer networks.
\newblock {\em J. Phys. C}, 32:335102(1--20), 2020.

\bibitem{haydukivskauniversal}
Khristine Haydukivska, Viktoria Blavatska, and Jaros{\l}aw Paturej.
\newblock Universal size ratios of gaussian polymers with complex architecture:
  radius of gyration vs hydrodynamic radius.
\newblock {\em Scientific Reports}, 10(1):14127, 2020.

\bibitem{Edwards}
S~F Edwards.
\newblock The statistical mechanics of polymers with excluded volume.
\newblock {\em Proceedings of the Physical Society}, 85(4):613, 1965.

\bibitem{grest1987}
Gary~S. Grest, Kurt Kremer, and T.~A. Witten.
\newblock Structure of many arm star polymers: a molecular dynamics simulation.
\newblock {\em Macromolecules}, 20(6):1376--1383, nov 1987.

\bibitem{lammps}
Steve Plimpton.
\newblock Fast parallel algorithms for short-range molecular dynamics.
\newblock {\em Journal of Computational Physics}, 117(1):1--19, 1995.

\bibitem{Madras88}
Neal Madras and Alan~D. Sokal.
\newblock The pivot algorithm: A highly efficient monte carlo method for the
  self-avoiding walk.
\newblock {\em J. Stat. Phys.}, 50(1-2):109--186, January 1988.

\bibitem{Clisby10}
Nathan Clisby.
\newblock Efficient implementation of the pivot algorithm for~self-avoiding
  walks.
\newblock {\em J. Stat. Phys.}, 140(2):349--392, May 2010.

\bibitem{Wei}
Gaoyuan Wei.
\newblock New approaches to shapes of arbitrary random walks.
\newblock {\em Physica A: Statistical Mechanics and its Applications},
  222(1):155--160, 1995.

\bibitem{Blavatska12}
V~Blavatska, C~von Ferber, and Yu~Holovatch.
\newblock Disorder effects on the static scattering function of star branched
  polymers.
\newblock {\em Condensed Matter Physics}, 15(3):33603, sep 2012.

\end{thebibliography}

 \end{document}